\documentclass[titlepage]{elsart}
\usepackage[dvips]{graphicx}
\usepackage{amssymb}
\usepackage[dvips]{epsfig}
\begin{document}

\begin{frontmatter}
\title{A Dirac Sea for a General Non-Inertial Observer in Flat \boldmath{1+1} Dimensional Spacetime}
\author{Mark D. Goodsell} \address{Hertford College, Catte Street, Oxford, OX1 3BW, UK}  \author{Carl E. Dolby} \address{Department of Theoretical Physics, 1 Keble Road, Oxford, OX1 3RH, UK} \author{Stephen F. Gull} \address{Astrophysics Group, Cavendish Laboratory, Madingley Road, Cambridge, CB3 0HE, UK}

\begin{abstract}
A coordinate system is set up for a general accelerating observer and is used to determine the particle content of the Dirac vacuum for that observer. Equations are obtained for the spatial distribution and total number of particles for massless fermions as seen by this observer, generalising previous work.
\end{abstract}

\end{frontmatter}

\section{Introduction}
Unruh pointed out in 1976 \cite{Un} that an observer who accelerated uniformly through the Minkowski vacuum would see a ``thermal" spectrum of scalar particles, whose apparent temperature depends upon the acceleration. The same result has also been found for fermions \cite{Soffel}. The constant acceleration case has been extensively investigated and there is a good review by S. Tagaki \cite{Takagi}. However, there have been few attempts at the general case.

The problem of dealing with an accelerated observer is partly how to construct a coordinate system which describes measurements made by that observer. In an inertial reference frame we can easily envisage such a system by, for example, imagining an infinite array of rods and clocks (see, for example, \cite{Taylor}). If the frame is accelerating we could consider momentarily co-moving inertial frames, but there is then an ambiguity in the coordinates assigned by the observer to events far away in space-time from the observer's world line. We are led to believe that there is no ``natural" coordinate system for the observer \cite{Misner2}. However, when considering observer-dependent and non-local concepts such as `vacuum' and `particle', it is necessary to ensure that the coordinates relate to the observer we are considering.

Our solution to this problem is based on the measurements that an observer could make. We shall use the ``radar time" coordinate construction for an arbitrary observer as defined by Dolby \cite{Dolbyradar}. We then apply this construction to the observer's concept of particle, and derive an expression for the number of fermions that this observer would find in the inertial vacuum. 

We use ``natural" units throughout such that $\hbar = c = 1 $. We also restrict our study to $1 + 1 $ (one time plus one space) dimensions.

\section{Radar Coordinates}
\subsection{General Derivation}

The concept of radar coordinates for an accelerated observer is not new, and an excellent list of references is given in \cite{Dolbyradar}. A set of coordinates is generated having one-to-one correspondence with the whole region of the Lorentz frame with which the accelerated observer can exchange signals. A good derivation, and proofs of various properties, is given also in \cite{Pauri}, in which the construction is referred to as ``M\"arzke-Wheeler coordinates".

Consider now an observer whose worldline $\lambda $ in an inertial frame is described by $x^\mu = x^{\mu }_{\lambda } (\tau_\lambda ) $ , where $\tau_\lambda $ is the observer's proper time and $x^\mu $ are the Minkowski coordinates. A signal is sent to the event whose coordinates are to be measured; the time at which the signal is sent is noted, and a signal is received from the event at a later time. The two times can be used to determine both the time and position of the event according to the observer. Assuming that the observer moves in only one dimension and does not rotate, we define:

\begin{quote}
\begin{tabular}{lp{9.8cm}} $\tau_1 \equiv $ & Time at which the signal travelling in the observer's backward direction is received or is sent.\\
$\tau_2 \equiv $ & Time at which the signal travelling in the observer's forward direction is received or is sent. \end{tabular}
\end{quote}

The `radar time' $\tau $ of the event is defined as the mean of these times, while the `radar distance' $\rho $ is one half of their difference (in natural units):

\begin{eqnarray}\tau \equiv \frac{1}{2}(\tau_1 + \tau_2) \nonumber \\
\label{eq:rt}\rho \equiv \frac{1}{2}(\tau_1 - \tau_2) \end{eqnarray}
We now define the variables $x^+ $ and $x^- $ :
\begin{eqnarray}x^+ \equiv t + z \nonumber \\
x^- \equiv t - z \label{eq:xpm}\end{eqnarray}
where $t$ and $z$ are the coordinates $x^0 $ and $x^1 $ in our Minkowski frame. It follows that $\tau_1$, $\tau_2$ are related to $x^+ $, $x^- $ by 
\begin{eqnarray}x^+ = x^{+}_{\lambda}(\tau_1) \nonumber \\ 
x^- = x^{-}_{\lambda}(\tau_2) \end{eqnarray}
since the signals follow paths of constant $x^- $ for the `forward' rays, and paths of constant $x^+ $ for the `backward' ones in a flat space. 

\begin{figure} 
\begin{center}
\epsfig{file=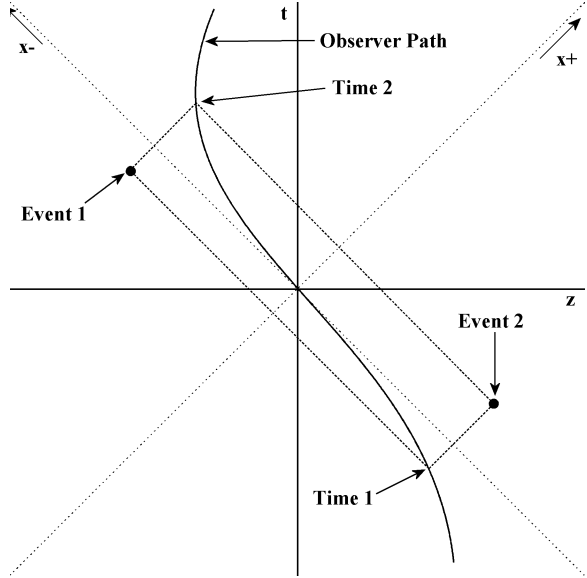,height=3.0in}
\caption{Radar Time for a general observer} \label{fig:radar1} 
\end{center}
\end{figure}

Figure \ref{fig:radar1} illustrates this construction. Times `1' and `2' correspond to $\tau_1 $ and $\tau_2 $ for `Event 1', and to $\tau_2 $ and $\tau_1 $ for `Event 2', so that the events are simultaneous according to the observer, but the values of $\rho $ have opposite sign. This convention differs slightly from that defined in \cite{Dolbyradar}; we have defined $\rho $ as a `spatial' rather than a `radial' coordinate so that our $\vert \rho \vert $ corresponds to $\rho $ of \cite{Dolbyradar} - this change is convenient for $1 + 1 $ dimensions.

\subsection{Metric}

We can use the coordinate transformation to calculate the metric for the observer, using the invariance of the space-time interval $ds $. In our Lorentz frame, we have:
\begin{equation}
ds^2 = dt^2 - dz^2 = dx^+ dx^- 
\end{equation}
However,
\begin{eqnarray}
dx^+ = dx^{+}_{\lambda} = \frac{dx^{+}_{\lambda}}{d\tau_1}d\tau_1 = \frac{dx^{+}_{\lambda}}
{d\tau_1}(d\tau + d\rho) \nonumber \\
dx^- = dx^{-}_{\lambda} = \frac{dx^{-}_{\lambda}}{d\tau_2}d\tau_2 = \frac{dx^{-}_{\lambda}}{d\tau_2}(d\tau - d\rho) 
\end{eqnarray}
This identity is one of several involving radar coordinates; others are given in the Appendix. By defining
\begin{equation}
U^+(\tau_1) = \frac{dx^{+}_{\lambda}(\tau_1)}{d\tau_1}, \qquad U^-(\tau_2) = \frac{dx^{-}_{\lambda}(\tau_2)}{d\tau_2} 
\end{equation}
we have: 
\begin{equation} \label{eqn:interval}
ds^2 = U^+ U^- (d\tau^2 - d\rho^2 )
\end{equation}
This expression gives the metric \textbf{g}, with components $g_{\mu \nu} $, via the equation
\begin{equation}
ds^2 = g_{\mu \nu} dx^\mu dx^\nu 
\end{equation}
where the Einstein summation convention is employed (as it will be throughout). In matrix form, the metric can be written:

\begin{equation}
\mathbf{g} = \left( \begin{array}{cc} U^+ U^- & 0 \\ 0 & - U^+ U^- \end{array} \right)
\label{metric} \end{equation}
The off-diagonal elements are zero since the coefficient of $dx^1 dx^0 $ in the interval is zero. 

If the observer's velocity has a discontinuity (i.e. there is a momentary infinite acceleration) then the metric will be two-valued at all events having the same value of $x^+ $ or $x^- $ as the observer at the point of the discontinuity - i.e. all along the future and past light cones stemming from that event. Also, on the observer's worldline, $\tau_1 = \tau_2 = \tau_\lambda$ so that $U^+ U^- = 1$ and $U^{\pm} = \sqrt{\frac{1\pm v}{1\mp v}}$, where $v(\tau_\lambda) $ is the speed of the observer.

\subsection{Examples}

\subsubsection{Inertial Frame}
\label{Inertial}

A very simple case that provides a useful consistency check is an inertial frame with a velocity $v $ relative to our original frame. From the Lorentz transformation, we have 
\begin{equation}
t = \gamma \tau_\lambda, \qquad z = \gamma v\tau_\lambda
\end{equation}
and
\begin{eqnarray}
x^+ = \gamma \tau_1 (1 + v) = \tau_1 \sqrt{\frac{1 + v}{1 - v}} \nonumber \\
x^- = \gamma \tau_2 (1 - v) = \tau_2 \sqrt{\frac{1 - v}{1 + v}}
\end{eqnarray}
giving
\begin{equation}
\tau = \gamma (t - vz), \qquad \rho = \gamma (z - vt)
\end{equation}
as expected. We also have
\begin{equation}
U^+ = \sqrt{\frac{1 + v}{1 - v}}, \qquad  U^- = \sqrt{\frac{1 - v}{1 + v}}
\end{equation}
which gives our Minkowski metric
\begin{equation}\mathbf{g} = \left( \begin{array}{cc} 1 & 0 \\ 0 & - 1 \end{array} \right) \end{equation}

\subsubsection{Constant Acceleration}
\label{constant}

The simplest nontrivial case is with constant acceleration. As set out in \cite{Misner2}, the observer's worldline can be written:
\begin{equation}
t = g^{-1} \sinh g\tau_\lambda , \qquad  z = g^{-1} \cosh g\tau_\lambda
\end{equation}
where $g$ is the acceleration. Hence
\begin{eqnarray}
x^+ = \frac{e^{g\tau_1}}{g}, \qquad  x^- = - \frac{e^{-g\tau_2}}{g} \nonumber \\
\tau = -\frac{\ln -\frac{x^+}{x^-}}{2g}, \qquad \rho = \frac{\ln -g^2 x^+ x^-}{2g}
\end{eqnarray}
and
\begin{equation}
U^+ = e^{g\tau_1}, \qquad  U^- = e^{-g\tau_2}
\end{equation}
giving
\begin{equation}
\mathbf{g} = \left( \begin{array}{cc} e^{2g\rho} & 0 \\ 0 & - e^{2g\rho} \end{array} \right)
\end{equation}

This result agrees with \cite{Dolbyradar} and \cite{Pauri}.  

\subsubsection{Constant Acceleration Beginning at t = 0}
\label{interesting}

\begin{figure} 
\begin{center}
\epsfig{file=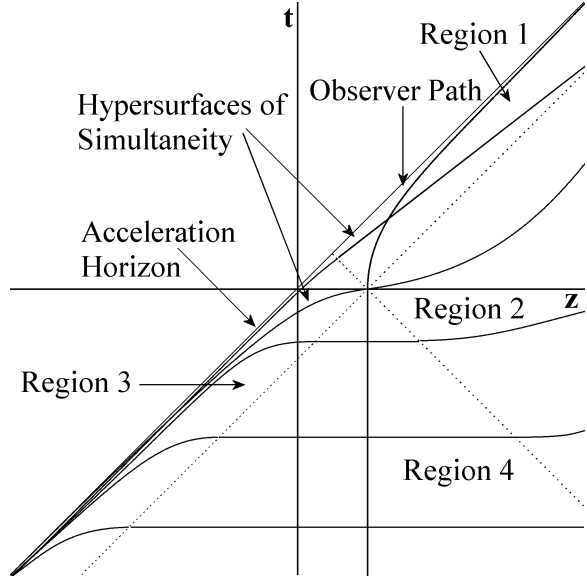,height=3.0in}
\caption{Radar time for an observer who accelerates constantly after $t = 0$} \label{fig:radar2} 
\end{center}
\end{figure}

In this case the observer's world line is described by: 
\begin{eqnarray}
z & = & \left \{ \begin{array}{ll} g^{-1} \cosh g\tau_\lambda \qquad & \tau_\lambda \ge 0 \\ g^{-1} & \tau_\lambda \le 0 \end{array} \right. \nonumber \\
t & = & \left \{ \begin{array}{ll} g^{-1} \sinh g\tau_\lambda \qquad & \tau_\lambda \ge 0 \\ \tau_\lambda & \tau_\lambda \le 0 \end{array} \right.
\end{eqnarray}
The radar coordinate system divides naturally into four regions as shown in Figure \ref{fig:radar2}; the corresponding coordinates are given in Table \ref{tab}. The hypersurfaces of simultaneity are not easy to calculate exactly in regions 2 and 3.

\begin{table}
\begin{tabular}{ccccccc}
Region & $x^+ $ & $x^- $ & $U^+ $ & $U^- $ & $\rho $ & $\tau $ \\ \hline
1 & $\frac{e^{g\tau_1}}{g} $ & $- \frac{e^{-g\tau_2}}{g} $ & $e^{g\tau_1} $ & $e^{-g\tau_2} $ & $\frac{\ln -g^2 x^+ x^-}{2g} $ & $-\frac{\ln -\frac{x^+}{x^-}}{2g} $ \\
2 & $\frac{e^{g\tau_1}}{g} $ & $\tau_2 - \frac{1}{g} $ & $e^{g\tau_1} $ & 1 & $\frac{1}{2} ( \frac{\ln gx^+}{g} - x^- - \frac{1}{g} ) $ & $\frac{1}{2} ( \frac{\ln gx^+}{g} + x^- + \frac{1}{g} ) $ \\ 
3 & $\tau_1 + \frac{1}{g} $ & $- \frac{e^{-g\tau_2}}{g} $ & 1 & $e^{-g\tau_2} $ & $\frac{1}{2} ( x^+ - \frac{1}{g} + \frac{\ln -gx^-}{g} ) $ & $\frac{1}{2} ( x^+ - \frac{1}{g} - \frac{\ln -gx^-}{g} ) $ \\
4 & $\tau_1 + \frac{1}{g} $ & $\tau_2 - \frac{1}{g} $ & 1 & 1 & z & t 
\end{tabular}
\caption{Radar time values for observer accelerated uniformly from $t = 0$}\label{tab}\end{table}

\subsubsection{Constant acceleration for a finite time}
\label{turnaround}

\begin{figure} 
\begin{center}
\epsfig{file=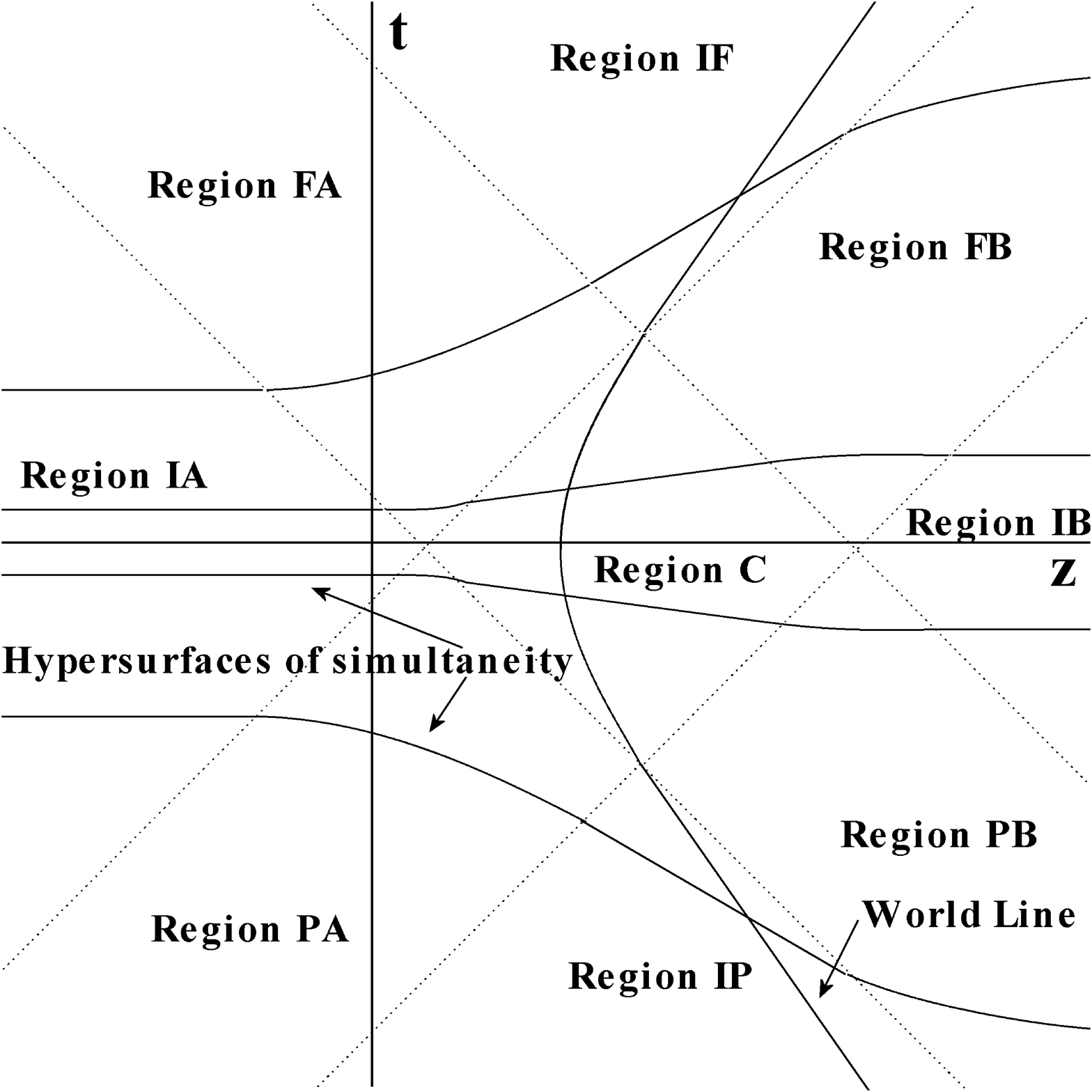,height=4.0in}
\caption{Worldline, radar coordinate regions and hypersurfaces of simultaneity for an observer who accelerates constantly for a finite time} \label{fig:radar3} 
\end{center}
\end{figure}

In this case an observer travels at constant speed until a certain time, then accelerates constantly until the speed is reversed, i.e. equal but in the opposite direction to the initial value; this speed is then maintained. The worldline is given by:

\begin{eqnarray}
z & = & \left \{ \begin{array}{ll} \gamma v(\tau_\lambda - \tau_c ) + g^{-1} \cosh g\tau_c \qquad & \tau_\lambda \ge \tau_c \\  g^{-1} \cosh g\tau_\lambda & \vert\tau_\lambda\vert \le \tau_c \\ - \gamma v (\tau_\lambda + \tau_c ) + g^{-1} \cosh g\tau_c & \tau_\lambda \le -\tau_c \\ \end{array} \right. \nonumber \\
t & = & \left \{ \begin{array}{ll} \gamma (\tau_\lambda - \tau_c ) + g^{-1} \sinh g\tau_c \qquad & \tau_\lambda \ge \tau_c \\ g^{-1} \sinh g\tau_\lambda & \vert\tau_\lambda\vert \le \tau_c \\ \gamma (\tau_\lambda + \tau_c ) - g^{-1} \sinh g\tau_c & \tau_\lambda \le -\tau_c \\ \end{array} \right.
\end{eqnarray}

where $\tau_c $ is the time at which acceleration ends according to the observer, and $v = \tanh g\tau_c$ is the final speed. Figure \ref{fig:radar3} shows that the radar coordinate construction for this observer divides space-time naturally into nine regions as a result of the two discontinuities in the acceleration, though the observer can make measurements of the whole space-time. Values of the radar coordinate quantities in each region are given in Table \ref{tab2}.

\begin{table}
\begin{tabular}{ccccc}
Region & $U^+ $ & $U^- $ & $\rho $ & $\tau $ \\ \hline
C & $e^{g\tau_1} $ & $e^{-g\tau_2} $ & $\frac{\ln -g^2x^+x^-}{2g} $ & $-\frac{\ln \frac{-x^+}{x^-}}{2g} $ \\
IF & $\sqrt{\frac{1+v}{1-v}} $ & $\sqrt{\frac{1-v}{1+v}}$ & $ \gamma(z - vt) - \frac{1}{g} $ & $\gamma(t - vz) + \tau_c $ \\
IP & $\sqrt{\frac{1-v}{1+v}} $ & $\sqrt{\frac{1+v}{1-v}} $ & $\gamma(z + vt) - \frac{1}{g} $ & $\gamma(t + vz) - \tau_c $ \\
IA & $\sqrt{\frac{1-v}{1+v}} $ & $\sqrt{\frac{1-v}{1+v}} $ & $z\sqrt{\frac{1+v}{1-v}} - \frac{1}{g} - \tau_c $ & $t\sqrt{\frac{1+v}{1-v}} $ \\
IB & $\sqrt{\frac{1+v}{1-v}} $ & $\sqrt{\frac{1+v}{1-v}} $ & $z\sqrt{\frac{1-v}{1+v}} + \tau_c - \frac{1}{g} $ & $t\sqrt{\frac{1-v}{1+v}} $ \\
FA & $e^{g\tau_1} $ & $\sqrt{\frac{1+v}{1-v}} $ & $\frac{1}{2}\left [ \frac{\ln gx^+}{g} - x^-\sqrt{\frac{1+v}{1-v}} - \frac{1}{g} - \tau_c \right ] $ & $\frac{1}{2}\left [ \frac{\ln gx^+}{g} + x^-\sqrt{\frac{1+v}{1-v}} + \frac{1}{g} + \tau_c \right ] $\\
FB & $\sqrt{\frac{1+v}{1-v}} $ & $e^{-g\tau_2} $ & $\frac{1}{2} \left [ \tau_c - \frac{1}{g} + x^+\sqrt{\frac{1-v}{1+v}} + \frac{\ln-gx^-}{g} \right ] $ & $\frac{1}{2} \left [ \tau_c - \frac{1}{g} + x^+\sqrt{\frac{1-v}{1+v}} - \frac{\ln-gx^-}{g} \right ] $ \\
PA & $\sqrt{\frac{1-v}{1+v}}$ & $e^{-g\tau_2} $ & $\frac{1}{2}\left [ x^+\sqrt{\frac{1-v}{1+v}} - \frac{1}{g} -\tau_c + \frac{\ln -gx^-}{g} \right ] $ & $\frac{1}{2}\left [ x^+\sqrt{\frac{1-v}{1+v}} - \frac{1}{g} -\tau_c - \frac{\ln -gx^-}{g} \right ] $ \\
PB & $e^{g\tau_1} $ & $\sqrt{\frac{1+v}{1-v}} $ & $\frac{1}{2} \left [ \frac{\ln gx^+}{g} - x^-\sqrt{\frac{1-v}{1+v}} - \frac{1}{g} +\tau_c \right ] $ & $\frac{1}{2} \left [ \frac{\ln gx^+}{g} + x^-\sqrt{\frac{1-v}{1+v}} + \frac{1}{g} -\tau_c \right ] $

\end{tabular}
\caption{Radar coordinate values for an observer who accelerates constantly for a finite time}
\label{tab2}
\end{table}

This example is treated since it does not include an acceleration horizon, yet still uses much of the mathematics of the constant acceleration case. 

\section{Quantum Field Variables for the Accelerated Observer}

\subsection{Dirac Equation}

We study the Dirac vacuum, and hence use the Dirac equation for free particles as generalised to curved space-times/accelerated observers \cite{QEDSF}: 
\begin{equation}\imath\gamma^\mu\left (\frac{\partial}{\partial x^\mu}+\Gamma_\mu\right )\psi-m\psi = 0\label{diraceqn}\end{equation}
where $\psi $ is a 2-component spinor (in $1 + 1 $ dimensions),  $m$ is the mass of the spin-$1/2$ particle, and the $\Gamma_\mu $ are the ``spin connection" matrices, given by:
\begin{equation}\Gamma_\mu = \frac{1}{4}\gamma_\nu\left (\frac{\partial\gamma^\nu}{\partial x^\mu}+\left \{\nu \atop \lambda\mu \right \}\gamma^\nu \right )=\frac{1}{4}\gamma_\nu \nabla_\mu \gamma^\nu \label{connection}\end{equation}
Here $\left \{\nu \atop \lambda\mu \right \} $ are the conventional ``Christoffel symbols'' or ``connection coefficients'' which form part of the covariant derivative $\nabla_\mu $ \cite{Misner}. 

In $1 + 1 $ dimensions the quantities $\gamma^\mu $ are $2\times 2$ component matrices which transform as (contravariant) vectors, and they can be chosen to be any matrices that satisfy
\begin{equation}
\gamma^\mu \gamma^\nu + \gamma^\nu \gamma^\mu = 2g^{\mu \nu} \mathbb{I}
\label{gammalgebra} \end{equation}
where $\mathbb{I} $ is the identity matrix. We shall denote the $\gamma $-matrices in the Lorentz frame as $\bar{\gamma}^\mu $, so that in the accelerated frame we can choose:
\begin{equation}
\gamma^\mu = \frac{\partial x^\mu}{\partial \bar{x}^\alpha}\bar{\gamma}^\alpha
\label{transgam} \end{equation}
according to the transformation law for vectors, where the $\{ x^\mu \} $ represent the coordinates in our observer's frame, and the $\{ \bar{x}^\alpha \} $ are coordinates in our inertial frame. In this case $\Gamma_\mu = 0 $ and so the Dirac equation (\ref{diraceqn}) then becomes
\begin{equation}
\imath\gamma^\mu \frac{\partial \psi}{\partial x^\mu} - m\psi = 0
\label{Diracsimple} \end{equation}
From equation (\ref{transgam}), we can write this as
\begin{equation}
\imath\frac{\partial x^\mu}{\partial \bar{x}^\alpha}\frac{\partial \bar{x}^\beta}{\partial x^\mu}\bar{\gamma}^\alpha  \frac{\partial \psi}{\partial \bar{x}^\beta} - m\psi = 0
\end{equation}
which simplifies to
\begin{equation}
\imath\bar{\gamma}^\alpha \frac{\partial \psi}{\partial \bar{x}^\alpha} - m\psi = 0
\end{equation}
and shows the covariance of the equation - the solutions will be the same in all frames. We can rewrite this using equation (\ref{gammalgebra}) as 
\begin{equation}
\frac{\partial \psi}{\partial \bar{x}^0} + \bar{\gamma}^0 \bar{\gamma}^1 \frac{\partial \psi}{\partial \bar{x}^1} = - \imath m\bar{\gamma}^0 \psi
\end{equation}

We choose the $\bar{\gamma} $-matrices to be
\begin{equation}
\bar{\gamma}^0 = \left ( \begin{array}{cc} 0 & 1 \\ 1 & 0 \end{array} \right ) , \qquad \bar{\gamma}^1 = \left ( \begin{array}{cc} 0 & -1 \\ 1 & 0 \end{array} \right )
\end{equation}
with the basis spinors
\begin{equation}
\phi_+ = \left ( 1 \atop 0 \right ) , \qquad \phi_- = \left ( 0 \atop 1 \right )
\end{equation}
which gives
\begin{equation}
\bar{\gamma}^0 \bar{\gamma}^1 = \left ( \begin{array}{cc} 1 & 0 \\ 0 & -1 \end{array} \right )
\end{equation}
We then write $\psi = \psi_+ \phi_+ + \psi_- \phi_- $ to obtain the coupled equations:
\begin{eqnarray}
\frac{\partial \psi_+}{\partial t} + \frac{\partial \psi_+}{\partial z} = -\imath m \psi_- \nonumber \\
\frac{\partial \psi_-}{\partial t} - \frac{\partial \psi_-}{\partial z} = -\imath m \psi_+
\end{eqnarray}
For massless fermions, these equations decouple to give 
\begin{equation}
\psi_+ = \psi_+ (x^-), \qquad \psi_- = \psi_- (x^+)
\label{massless} \end{equation}
whereas for the massive case the equations become
\begin{eqnarray}
\frac{\partial^2 \psi_+ }{\partial t^2 } - \frac{\partial^2 \psi_+ }{\partial z^2 } = -m^2 \psi_+ \nonumber \\
\frac{\partial^2 \psi_- }{\partial t^2 } - \frac{\partial^2 \psi_- }{\partial z^2 } = -m^2 \psi_-
\end{eqnarray}
which can be solved by separation of variables to give:
\begin{equation}
\psi_+ = N e^{\imath (At + Bz)}, \qquad \psi_- = - \frac{A + B}{m} \psi_+
\end{equation}
where $N $ is a suitable normalisation, and $A $ and $B $ are constants obeying $A^2 - B^2 = m^2 $. In our inertial frame, $A $ and $B $ respectively represent the energy and momentum of the particle, although they are combined as components of the momentum-energy vector between inertial frames. This identification does not hold for our accelerated frame, in which the energy and momentum operators are different. To determine the energy and to identify wavefunctions as representing particles or antiparticles we must calculate the appropriate Hamiltonian \cite{Dolby3}. As a preliminary step we evaluate the inner product in our coordinate system.

\subsection{Inner Product}

The generalised inner product of two spinors $\psi $ and $\phi $ in an accelerated frame on a hypersurface $\Sigma $ of constant $\tau $ is:
\begin{equation}
\langle \psi \vert \phi \rangle_\Sigma = \int\limits_{\Sigma} d\Sigma_\mu \sqrt{- g} \bar{\psi}\gamma^\mu \phi 
\end{equation}
where $g $ is the determinant of $g_{\mu \nu} $, and $\bar{\psi} $ is the conjugate of $\psi $, defined as $\bar{\psi} \equiv \psi^\dagger \bar{\gamma}^0 $. The `surface' element $d\Sigma $ is given in $1 + 1 $ dimensions by
\begin{equation}
d\Sigma_\mu = \frac{\frac{\partial \tau}{\partial x^\mu}}{\frac{\partial \tau}{\partial x^0}}dx^1
\end{equation}
The above definition of the inner product is independent of the choice of $\Sigma $, provided that $\psi $ and $\phi $ are solutions of the Dirac equation. 

Since $\frac{\partial \tau}{\partial \rho} = 0 $, we can write the inner product in radar coordinates as
\begin{equation}
\langle \psi \vert \phi \rangle_\tau = \int\limits_{- \infty}^{\infty} d\rho \sqrt{- g}\psi^\dagger \bar{\gamma}^0 \gamma^0 \phi 
\end{equation}
Equations (\ref{metric}) and (\ref{transgam}) allow us to express this as
\begin{equation}
\langle \psi \vert \phi \rangle_\tau = \int\limits_{- \infty}^{\infty} d\rho \, U^+ U^- \psi^\dagger \bar{\gamma}^0 \left ( \frac{\partial \tau}{\partial t}\bar{\gamma}^0 + \frac{\partial \tau}{\partial z}\bar{\gamma}^1 \right ) \phi 
\end{equation}
which finally becomes
\begin{equation}
\langle \psi \vert \phi \rangle_\tau = \frac{1}{2}\int\limits_{- \infty}^{\infty} d\rho \: \psi^\dagger \left [ U^- (\mathbb{I} + \bar{\gamma}^0 \bar{\gamma}^1 ) + U^+ (\mathbb{I} - \bar{\gamma}^0 \bar{\gamma}^1 ) \right ] \phi 
\label{inner1} \end{equation}

Alternatively, the inner product (still evaluated on a hypersurface of constant $\tau $ ) can be written in inertial coordinates as:

\begin{equation}
\langle \psi \vert \phi \rangle_\tau = \int\limits_{\Sigma} dz \,\psi^\dagger \left ( \mathbb{I} + \frac{\frac{\partial \tau}{\partial z}}{\frac{\partial \tau}{\partial t}} \bar{\gamma}^0 \bar{\gamma}^1 \right ) \phi 
\end{equation}

\subsection{Hamiltonian}
\label{Hamiltonian}

We can rewrite equation (\ref{Diracsimple}) as 
\begin{equation}
\imath \frac{\partial \psi}{\partial \tau} = - \frac{\gamma^0}{g^{00}} \left ( \imath \gamma^1 \frac{\partial \psi}{\partial \rho} - m\psi \right )
\end{equation}
and so define the Hamiltonian $\hat{H}_{nh} $ as
\begin{equation}
\hat{H}_{nh} = - U^+ U^- \gamma^0 \left ( \imath \gamma^1 \frac{\partial }{\partial \rho} - m \right )
\end{equation}
Now,
\begin{eqnarray}
\gamma^0 \gamma^1 & = & \frac{\partial \tau}{\partial \bar{x}^\alpha} \frac{\partial \rho}{\partial \bar{x}^\beta} \bar{\gamma}^\alpha \bar{\gamma}^\beta \nonumber \\
& = & \left ( \frac{\partial \tau}{\partial t} \frac{\partial \rho}{\partial t} - \frac{\partial \tau}{\partial z} \frac{\partial \rho}{\partial z} \right ) \mathbb{I}  +  \left ( \frac{\partial \tau}{\partial t} \frac{\partial \rho}{\partial z} - \frac{\partial \tau}{\partial z} \frac{\partial \rho}{\partial t} \right ) \bar{\gamma}^0 \bar{\gamma}^1 \nonumber \\
& = & g^{01} \mathbb{I}  +  \left ( \left ( \frac{\partial \tau}{\partial t} \right )^2 - \left ( \frac{\partial \tau}{\partial z}\right )^2 \right ) \bar{\gamma}^0 \bar{\gamma}^1 \nonumber \\
& = & g^{00} \bar{\gamma}^0 \bar{\gamma}^1
\end{eqnarray}
and
\begin{eqnarray}
\gamma^0 & = & \left ( \frac{\partial \tau}{\partial x^+} + \frac{\partial \tau}{\partial x^-} \right ) \bar{\gamma}^0 + \left ( \frac{\partial \tau}{\partial x^+} - \frac{\partial \tau}{\partial x^-} \right ) \bar{\gamma}^1 \nonumber \\
& = & \frac{1}{2} \left [ \left ( \frac{1}{U^+} + \frac{1}{U^-} \right ) \bar{\gamma}^0 + \left ( \frac{1}{U^+} - \frac{1}{U^-} \right ) \bar{\gamma}^1 \right ]
\end{eqnarray}
so that
\begin{equation}
\hat{H}_{nh} = - \imath \bar{\gamma}^0 \bar{\gamma}^1 \frac{\partial }{\partial \rho} + \frac{m}{2} \left [ U^+ \left ( \bar{\gamma}^0 - \bar{\gamma}^1 \right ) +U^- \left ( \bar{\gamma}^0 + \bar{\gamma}^1 \right ) \right ]
\label{Hnh} \end{equation}
where we have used some of the identities given in the Appendix. 

It is clear from equation (\ref{inner1}) that the Hamiltonian in (\ref{Hnh}) is not Hermitian - hence the subscript `nh'. The explanation lies in the time dependence of both the inner product and the Hamiltonian, although $\hat{H}_{nh} $ is time-independent in the case of massless fermions. At first sight this would appear to violate unitarity. However, as stated by Dolby \emph{et al.}\cite{Dolby}, this is not so; as a result of the time-dependence in the inner product volume element, unitarity actually requires a non-Hermitian Hamiltonian. Dolby \emph{et al.} \cite{Dolby} use the stress-energy tensor to evaluate the energy of the state, and hence find the Hamiltonian at a certain time $\tau_0 $. We quote the result that the Hamiltonian $\hat{H}_1 $ for states at a time $\tau_0 $ can be written as
\begin{equation}
\hat{H}_1 \equiv \frac{1}{2} \left \{ \hat{H}_{nh} + \hat{H}_{nh}^{\dagger} \right \}
\end{equation}
The eigenstates of this Hamiltonian can be classified as positive or negative frequency according to whether their eigenvalues are positive or negative. 

To find the Hermitian conjugate of our Hamiltonian, we use two arbitrary spinors $\psi $ and $\phi $ together with the definition
\begin{equation}
\langle \hat{H}_{nh}\psi\vert \phi \rangle = \langle \psi\vert\hat{H}_{nh}^{\dagger} \phi \rangle
\end{equation}
We therefore consider
\begin{equation}
\langle \hat{H}_{nh}\psi\vert \phi \rangle_{\Sigma_{\tau_0}} = \frac{1}{2} \int\limits_{-\infty}^{\infty} \! d\rho \left ( \hat{H}_{nh}\psi \right )^{\dagger}  \left [ U^- (\mathbb{I} + \bar{\gamma}^0 \bar{\gamma}^1 ) + U^+ (\mathbb{I} - \bar{\gamma}^0 \bar{\gamma}^1 ) \right ] \phi 
\end{equation}
Define $\sigma = \bar{\gamma}^0 \bar{\gamma}^1 $, $a_1 = \bar{\gamma}^0 + \bar{\gamma}^1 $, $a_2 = \bar{\gamma}^0 - \bar{\gamma}^1 $, $b_1 = \mathbb{I} + \bar{\gamma}^0 \bar{\gamma}^1 $ and $b_2 = \mathbb{I} - \bar{\gamma}^0 \bar{\gamma}^1 $, so that
\begin{equation}
a_1^\dagger = a_2, \qquad b_1^{\dagger} = b_1, \qquad b_2^\dagger = b_2, \qquad \sigma^\dagger = \sigma
\end{equation}
with further identities listed in the appendix. If we now put $M = \frac{1}{2} \left [ U^- b_1 + U^+ b_2 \right ] $, then 
\begin{equation}
M^{-1} = \frac{1}{2U^+ U^-} \left ( U^+ b_1 + U^- b_2 \right )
\end{equation}
By using these definitions, we can write
\begin{eqnarray}
\langle \hat{H}_{nh}\psi\vert \phi \rangle_{\Sigma_{\tau_0}} & = & \int\limits_{-\infty}^{\infty} d\rho \left ( -\imath \sigma \frac{\partial \psi}{\partial \rho} + \frac{m}{2} \left [ U^+ a_2 + U^- a_1 \right ]\psi \right )^\dagger M \phi \nonumber \\
& = & \int\limits_{-\infty}^{\infty} d\rho \left ( \imath  \frac{\partial \psi^{\dagger}}{\partial \rho} \sigma + \frac{m\psi^{\dagger}}{2} \left [ U^+ a_1 + U^- a_2 \right ] \right ) M \phi  \nonumber \\
& = & \left [ \psi^\dagger ( \imath\sigma ) M\phi \right ]_{-\infty}^{\infty} \nonumber \\
&   & - \int\limits_{-\infty}^{\infty} d\rho \: \psi^{\dagger} \imath\sigma \frac{\partial}{\partial \rho}  \left ( M\phi \right )  \nonumber \\
&   & + \frac{m}{2} \int\limits_{-\infty}^{\infty} d\rho \: \psi^{\dagger} M \left ( M^{-1}\left [ U^+ a_1 + U^- a_2 \right ] M \right ) \phi  \nonumber \\
& = & \int\limits_{-\infty}^{\infty} d\rho \: \psi^{\dagger} M ( -\imath\sigma \frac{\partial \phi}{\partial \rho} )  \nonumber \\
&   & - \int\limits_{-\infty}^{\infty} d\rho \: \psi^{\dagger} ( \imath \sigma \frac{\partial M}{\partial \rho} ) \phi  \nonumber \\
&   & + \frac{m}{2} \int\limits_{-\infty}^{\infty} d\rho \: \psi^\dagger M  \left [ U^+ a_2 + U^- a_1 \right ]  \phi  \nonumber \\
& = & \langle \psi\vert -\imath\sigma \frac{\partial}{\partial \rho}\vert \phi \rangle_{\Sigma_{\tau_0}} \nonumber \\
&   & - \int\limits_{-\infty}^{\infty} d\rho \: \psi^\dagger  M \left (M^{-1} \imath\sigma  \frac{\partial M}{\partial \rho} \right ) \phi  \nonumber \\
&   & + \langle \psi\vert \frac{m}{2} \left [ U^+ a_2 + U^- a_1 \right ]\vert \phi \rangle_{\Sigma_{\tau_0}} 
\end{eqnarray}

We have applied the condition that the wavefunction must vanish at infinity, and have assumed that the functions $U^+ $ and $U^- $ are continuous. By applying further identities from the Appendix, we obtain
\begin{eqnarray}
\hat{H}_{nh}^{\dagger} & = & -i\sigma \frac{\partial}{\partial \rho} + \frac{\imath}{2}\left [ \frac{1}{U^-} \frac{dU^-}{d\tau_2} b_1 + \frac{1}{U^+} \frac{dU^+}{d\tau_1} b_2 \right ] \nonumber \\
&   & +  \frac{m}{2} \left [ U^+ a_2 + U^- a_1 \right ]
\end{eqnarray}
This finally gives us our Hermitian Hamiltonian $\hat{H}_1 $ as
\begin{eqnarray}
\hat{H}_1 & = & -i\sigma \frac{\partial}{\partial \rho} + \frac{\imath}{4}\left [ \frac{1}{U^-} \frac{dU^-}{d\tau_2} b_1 + \frac{1}{U^+} \frac{dU^+}{d\tau_1} b_2 \right ] \nonumber \\
&   & +  \frac{m}{2} \left [ U^+ a_2 + U^- a_1 \right ]
\label{ham} \end{eqnarray}

We can use this Hamiltonian to calculate the positive and negative eigenstates of the Dirac vacuum, as measured by our accelerated observer at a certain time. These states do not cover the whole of space-time if an acceleration horizon exists, i.e. the observer approaches the speed of light and does not decelerate, so that signals from certain regions of space-time never reach the observer. However, as found in \cite{Dolby3}, the states are still sufficient to calculate the number of particles observed.

We call the eigenvalue of the Hamiltonian $\omega $, and write the eigenstates as $\psi_\tau = f_+ \phi_+ + f_- \phi_- $. Equation (\ref{ham}) then gives
\begin{eqnarray}
\omega f_+ & = & - \imath \frac{\partial f_+ }{\partial \rho }  +  \frac{\imath }{2} \frac{1}{U^- } \frac{dU^- }{ d \tau_2 } f_+ + mU^+ f_- \nonumber \\
\omega f_- & = & \imath \frac{\partial f_-}{\partial \rho }  +  \frac{\imath}{2} \frac{1}{U^+ } \frac{d U^+ }{ d \tau_1 } f_- + mU^- f_+
\end{eqnarray}
We shall not attempt to solve these coupled equations here, but consider only the massless case.

\subsubsection{Eigenstates for Massless Fermions}

For massless fermions, our Hamiltonian reduces to
\begin{equation}
\hat{H}_1 = -\imath\sigma \frac{\partial}{\partial \rho} + \frac{\imath}{4}\left [ \frac{1}{U^-} \frac{dU^-}{d\tau_2} b_1 + \frac{1}{U^+} \frac{dU^+}{d\tau_1} b_2 \right ]
\end{equation}
for which we require a complete set of eigenstates. If we consider a time $\tau_0 $, we can use equation (\ref{rhoparams}) from the Appendix to obtain 
\begin{eqnarray}
\hat{H}_1 & = & -\imath\sigma \frac{\partial}{\partial \tau_1} + \frac{\imath}{4}\left [ \frac{1}{U^-} \frac{dU^-}{d\tau_2} b_1 + \frac{1}{U^+} \frac{dU^+}{d\tau_1} b_2 \right ] \nonumber \\
& = & \imath\sigma \frac{\partial}{\partial \tau_2} + \frac{\imath}{4}\left [ \frac{1}{U^-} \frac{dU^-}{d\tau_2} b_1 + \frac{1}{U^+} \frac{dU^+}{d\tau_1} b_2 \right ]
\end{eqnarray}
which gives 
\begin{eqnarray}
\omega f_+ & = & \imath \frac{df_+ }{d\tau_2 }  +  \frac{\imath }{2} \frac{1}{U^- } \frac{dU^- }{ d \tau_2 } f_+ \nonumber \\
\omega f_- & = & \imath \frac{df_-}{d\tau_1 }  +  \frac{\imath}{2} \frac{1}{U^+ } \frac{d U^+ }{ d \tau_1 } f_- 
\label{mlesseqns} \end{eqnarray}
A complete set of solutions can be found with $f_+ = f_+ ( \tau_2 ) $ and $ f_- = f_- ( \tau_1 ) $. That is, we can find solutions of the massless Dirac equation - given in equation (\ref{massless}) - that are  eigenstates at all times.  In consequence, the total number of massless fermions seen by a non-inertial observer is independent of time. The possibility of this is hinted at by the commutation of $\hat{H}_1 $ with the Dirac Operator $\imath \left ( \frac{\partial}{\partial \tau} + \sigma \frac{\partial}{\partial \rho} \right ) $. It occurs because the explicit time dependence disappears as a result of the decoupling of the equations, in contrast to the massive case. This result will be discussed in Section \ref{masslessfermions}.  

Equations (\ref{mlesseqns}) are simply solved by use of an integrating factor to give
\begin{eqnarray}
f_+ & = & \frac{A_\omega e^{-\imath\omega\tau_2 }}{\sqrt{U^- }} \nonumber \\
f_- & = & \frac{B_\omega e^{-\imath\omega\tau_1 }}{\sqrt{U^+ }}
\end{eqnarray}
where $A_\omega$ and $B_\omega $ are normalisation constants. If we normalise the wavefunctions to  $2\pi \delta (\omega - \omega^\prime ) $ we obtain:
\begin{eqnarray}
\langle \psi_\omega \vert \psi_{\omega^\prime} \rangle & = & \int\limits_{-\infty}^{\infty} d\rho \left [ \frac{A_{\omega}^{*} e^{\imath\omega\tau_2}}{\sqrt{U^-}}\phi_{+}^{\dagger} + \frac{B_{\omega}^{*} e^{\imath\omega\tau_1}}{\sqrt{U^+}}\phi_{-}^{\dagger} \right ] M \left [ \frac{A_{\omega^\prime} e^{-\imath\omega^\prime \tau_2}}{\sqrt{U^-}}\phi_+ + \frac{B_{\omega} e^{-\imath\omega^\prime \tau_1}}{\sqrt{U^+}}\phi_- \right ]  \nonumber \\
& = & \int\limits_{-\infty}^{\infty} d\rho \left [ A_{\omega}^{*}A_{\omega^\prime} e^{\imath (\omega - \omega^\prime)\tau_2} \right ] + \left [ B_{\omega}^{*}B_{\omega^\prime} e^{\imath (\omega - \omega^\prime)\tau_1} \right ]  \nonumber \\
& = & 2\pi \delta (\omega - \omega^\prime ) \left ( A_{\omega}^{*}A_{\omega^\prime} + B_{\omega}^{*}
B_{\omega^\prime} \right ) \nonumber \\
& = & 2\pi \delta (\omega - \omega^\prime ) \left ( A_{\omega}^{*}A_{\omega} + B_{\omega}^{*}B_{\omega} \right )
\end{eqnarray}
Hence
\begin{equation}
A_{\omega}^{*}A_{\omega} + B_{\omega}^{*}B_{\omega} = 1
\end{equation}
and the coefficients $A_\omega $, $B_\omega $ are independent of $\omega $. We choose the ``forward-moving'' mode $\psi_{\omega\!,\, F} $ to have $\{A = 1 , B = 0 \}$ and the `backward-moving'' mode $\psi_{\omega\!,\, B} $ to have $\{A = 0 , B = 1 \} $.  

We now have a set of energy eigenstates which remain eigenstates at all times for our observer. We can classify the modes as positive or negative energy simply by the sign of $\omega $ in equation (\ref{mlesseqns}). We can also use these modes to calculate the particle content of the vacuum.

\section{Particle Content of the Dirac Vacuum}
\subsection{General Formulae}
\label{genpart}

According to the approach developed in \cite{Dolby3}, the particle density of the Dirac vacuum at a given time is given by
\begin{eqnarray}
n^+_\tau (x^1) = \sum_{\omega \ \omega^\prime}  \left ( (\beta\beta^\dagger )_{\omega\omega^\prime} \sqrt{-g} \psi_{\omega^\prime\!\!, +}^{\dagger} \bar{\gamma}^0 \gamma^\mu \psi_{\omega\!, +} \frac{\frac{\partial \tau}{\partial x^\mu}}{\frac{\partial \tau}{\partial x^0}} \right )  \nonumber \\
n^-_\tau (x^1) = \sum_{\omega \ \omega^\prime} \left ( (\gamma\gamma^\dagger )_{\omega\omega^\prime} \sqrt{-g} \psi_{\omega^\prime\!\!, -}^{\dagger} \bar{\gamma}^0 \gamma^\mu \psi_{\omega\!, -} \frac{\frac{\partial \tau}{\partial x^\mu}}{\frac{\partial \tau}{\partial x^0}} \right )
\label{density}\end{eqnarray}
where $n^+ $ and $n^- $ are the densities of particles and antiparticles, respectively defined such that $n^+_\tau (x^1)dx^1 $ is the number of particles in the length $dx^1 $ at $x^1$ (and similarly for $n^-_\tau (x^1) $ ), $\psi_{\omega, \pm} $ are the positive/negative energy eigenstates, and the summation implies integration over the variables between zero and infinity, with a division by $2\pi $ for each (due to normalisation), and the matrices $\beta $ and $\gamma $ are given by:
\begin{eqnarray}
\beta_{\omega p} = \langle \psi_{\omega\!, + } \vert u_{p, -} \rangle \nonumber \\
\gamma_{\omega p} = \langle \psi_{\omega\!, - } \vert u_{p, +} \rangle
\end{eqnarray}
where $u_{p, \pm } $ are the positive/negative energy Minkowski eigenstates with momentum $p$. These expressions lead to a total particle number $N^+ $ and antiparticle number $N^- $ in the observer's entire space as
\begin{eqnarray}
N^+ & = & \sum_{\omega \ \omega^\prime}   (\beta\beta^\dagger )_{\omega\omega^\prime } \langle \psi_{\omega^\prime\!\!, +} \vert \psi_{\omega\!, +} \rangle  \nonumber \\
& = & \ \int\limits_0^{\infty} d\omega (\beta\beta^\dagger )_{\omega\omega}  \nonumber \\
N^- & = & \sum_{\omega \  \omega^\prime}  (\gamma\gamma^\dagger )_{\omega\omega^\prime} \langle \psi_{\omega^\prime\!\!, -} \vert \psi_{\omega\!, -} \rangle  \nonumber \\
& = &  \ \int\limits_0^{\infty} d\omega (\gamma\gamma^\dagger )_{\omega\omega} 
\label{totalps} \end{eqnarray}
where we have used the orthogonality and normalisation properties of the eigenstates. We see that the eigenstates over the whole space-time are not required, just those for which our observer can construct coordinates.

\subsection{Massless Fermions}
\label{masslessfermions}

In the massless case, the Minkowski states are the plane-wave basis:
\begin{equation}
\begin{array}{lcll}
u_+  & = & e^{-\imath p x^- }\phi_+ \ \textrm{and } e^{-\imath p x^+ }\phi_- \qquad  & \textrm{for } p > 0  \\
u_-  & = & e^{\imath p x^- }\phi_+ \ \textrm{and } e^{\imath p x^+ }\phi_- \qquad  & \textrm{for } p > 0
\end{array}\end{equation}
where the $\phi_+ $ and $\phi_- $ components combine in the same way as the components of the eigenstates of our massless Hamiltonian, and are normalised to $2\pi \delta (p^\prime - p ) $. Taking inner products with our eigenstates of $\hat{H}_1 $, we must consider both ``forward-moving'' and ``backward-moving' particles, and must therefore evaluate the expressions for both of $\psi_{\omega\!,\, F} $ and $\psi_{\omega\!,\, B} $. The two sets need not have the same distribution or number of particles, merely the same number of  particles as antiparticles. Upon denoting ``forward-moving'' quantities with the subscript `F', and ``backward'' ones with a `B', we have:
\begin{eqnarray}
\beta_{\omega p, F} =\int\limits_{-\infty}^{\infty} d\tau_2 \: \sqrt{U^-} e^{\imath\omega\tau_2} e^{\imath p x^- }  \nonumber \\
\beta_{\omega p, B} =\int\limits_{-\infty}^{\infty} d\tau_1 \: \sqrt{U^+} e^{\imath\omega\tau_1} e^{\imath p x^+ } 
\label{betas} \end{eqnarray}
The quantity $(\beta\beta^\dagger )_{\omega\omega} $ gives the energy distribution of the particles observed. It is independent of time since the arguments in the inner product in $\beta_{\omega\! p} $ are both solutions of the Dirac equation. 

The equations for the antiparticles are exactly the same:
\begin{eqnarray}
\gamma_{\omega p, F} & = & \int\limits_{-\infty}^{\infty} d\rho \: \sqrt{U^-} e^{-\imath\omega\tau_2} e^{-\imath p x^- } \nonumber \\
& = & \beta_{\omega p, F}^{*} 
\end{eqnarray}
Similarly $\gamma_{\omega p, B} = \beta_{\omega p, B}^{*} $, so that
\begin{equation}
(\gamma\gamma^\dagger )_{\omega\omega^\prime\!\!, \, F}  =  (\beta\beta^\dagger )_{\omega\omega^\prime\!\!,\, F}^{*}, \qquad (\gamma\gamma^\dagger )_{\omega\omega^\prime\!\!,\, B}  =  (\beta\beta^\dagger )_{\omega\omega^\prime\!\!,\, B}^{*}
\end{equation}
The elements $(\gamma\gamma^\dagger )_{\omega\omega} $ and $(\beta\beta^\dagger )_{\omega\omega} $ are hence equal to each other, since the matrices are Hermitian, so from (\ref{totalps})the total number of `forward'-moving particles equals the total of `forward'-moving antiparticles, and similarly for `backward'-moving ones.

In this massless case, the formula for the distribution of particles also takes a simple form:
\begin{eqnarray}
n^+_F (\tau, \rho) & = &  \int\limits_{0}^{\infty}\frac{d\omega}{2\pi}\int\limits_{0}^{\infty}\frac{d\omega^\prime}{2\pi}(\beta\beta^\dagger )_{\omega\omega^\prime\!\!,\, F} \ e^{\imath (\omega^\prime - \omega )\tau_2}  \nonumber \\
n^-_F (\tau, \rho) & = &  \int\limits_{0}^{\infty}\frac{d\omega}{2\pi}\int\limits_{0}^{\infty}\frac{d\omega^\prime}{2\pi}(\gamma\gamma^\dagger )_{\omega\omega^\prime\!\!,\, F} \ e^{- \imath (\omega^\prime - \omega )\tau_2}  \nonumber \\
& = & \int\limits_{0}^{\infty}\frac{d\omega}{2\pi}\int\limits_{0}^{\infty}\frac{d\omega^\prime}{2\pi}(\beta\beta^\dagger )_{\omega^\prime\!\omega\!,\, F} \ e^{-\imath (\omega^\prime - \omega )\tau_2}  \nonumber \\
& = & n^+_F (\tau, \rho)
\end{eqnarray}
and similarly
\begin{equation}
n^\pm_B (\tau, \rho)  =   \int\limits_{0}^{\infty}\frac{d\omega}{2\pi}\int\limits_{0}^{\infty}\frac{d\omega^\prime}{2\pi}(\beta\beta^\dagger )_{\omega\omega^\prime\!\!,\, B} \ e^{\imath (\omega^\prime - \omega )\tau_1} 
\end{equation}
Therefore the distributions of particles and antiparticles are the same - as we expect for local conservation of charge (or lepton number etc). Since $ (\gamma\gamma^\dagger )_{\omega\omega^\prime} $ and $(\beta\beta^\dagger )_{\omega\omega^\prime} $ are independent of time, the particle distributions depend only on $\tau_1 $ for `backward'-moving particles, and $\tau_2 $ for the `forward'-movers.

Although the distributions of `forward'- and `backward'-moving particles are not in general the same, we expect them to be identical when the observer's trajectory is time-symmetric. To show this, consider integrating the expressions (\ref{betas}) along the hypersurface $\tau = 0 $ (this choice is merely for convenience, since we are free to choose any hypersurface of constant $\tau $). Time-symmetry implies that
\begin{eqnarray}
z_\lambda (-\tau_\lambda ) & = & z_\lambda (\tau_\lambda ) \nonumber \\
t_\lambda (-\tau_\lambda ) & = & - t_\lambda (\tau_\lambda ) 
\end{eqnarray}
and hence 
\begin{equation}
x^+_\lambda (\tau_\lambda ) = - x^-_\lambda (- \tau_\lambda ) 
\end{equation}
Along $\tau = 0 $ this gives 
\begin{eqnarray}
x^+ (\tau_1 ) = - x^- (-\tau_1 ) = - x^- (\tau_2) \nonumber \\
x^- (\tau_2 ) = -x^+ (-\tau_2 ) = -x^+ (\tau_1)  
\end{eqnarray}
and hence $U^+ = U^- $. Expression (\ref{betas}) for $\beta_{\omega \!p,\, B} $ thus becomes
\begin{eqnarray}
\beta_{\omega \!p,\, B} & = &\int\limits_{-\infty}^{\infty}d\tau_2 \: \sqrt{U^-} e^{-\imath\omega\tau_2} e^{-\imath p x^- (\tau_2) }  \nonumber \\
& = & (\beta_{\omega \!p,\, F})^*
\end{eqnarray}
The `forward'- and `backward' expressions for $(\beta\beta^\dagger )_{\omega\omega^\prime} $ must therefore be complex conjugates of each other. The total number of `forward'-moving particles therefore equals the number of `backward'-moving particles for a time-symmetric observer, and the frequency distributions must be the same. However, the spatial densities will be identical only at $\tau = 0 $, since the expressions in equation (\ref{density}) contain $\tau_2 $ for the `forward'-moving particles and $\tau_1 $ for the `backward'-moving particles.

\subsection{Examples}

\subsubsection{Inertial Frame}
\label{init}

By applying the results derived for massless fermions to a boosted Lorentz frame as in section \ref{Inertial}, we find that
\begin{eqnarray}
\beta_{\omega \!p,\, F} & = & \int\limits_{-\infty}^{\infty}d\rho \left ( \frac{1 - v}{1 + v} \right )^{\frac{1}{4}}  e^{\imath\omega\tau_2} e^{\imath p \tau_2 \sqrt{\frac{1 - v}{1 + v}} }  \nonumber \\
& = & \left ( \frac{1 - v}{1 + v} \right )^{\frac{1}{4}} \int\limits_{-\infty}^{\infty}d\rho \: e^{\imath\tau_2 \left ( \omega + \sqrt{\frac{1 - v}{1 + v}} p \right )} \nonumber \\
& = & 2\pi \left ( \frac{1 - v}{1 + v} \right )^{\frac{1}{4}} \delta(\omega + \sqrt{\frac{1 - v}{1 + v}} p ) \nonumber \\
& = & 0
\end{eqnarray}
The integral is zero, as expected, and in fact shows that a negative energy mode of energy $p $ is interpreted as a negative energy mode of energy $-\sqrt{\frac{1 - v}{1 + v}} p $. This is the correct Lorentz transformation for energy and momentum for a massless particle. 

We obtain the same result if we consider the $\phi_- $ components, or the antiparticle content.

\subsubsection{Constant Acceleration}

For massless fermions and a constantly accelerated observer, we obtain 
\begin{eqnarray}
\beta_{\omega \!p,\, F} & = & - \int\limits_{\infty}^{-\infty} d\tau_2 \: e^{(\imath\omega - \frac{g}{2} )\tau_2} e^{\frac{-\imath p e^{-g\tau_2}}{g}}  \nonumber \\
& = & \int\limits_{0}^{\infty}du \: (gu)^{\frac{-\imath\omega}{g} - \frac{1}{2} } e^{-\imath p u }  \qquad \textrm{where} \ u = \frac{e^{-g\tau_2}}{g} \nonumber \\
& = & \frac{1}{g} \left ( \frac{p}{g} e^{\frac{\imath \pi}{2}} \right )^{\frac{\imath\omega}{g} - \frac{1}{2} } \Gamma\left( \frac{-\imath\omega}{g} + \frac{1}{2} \right )
\end{eqnarray}
and similarly
\begin{eqnarray}
\beta_{\omega \!p,\, B} & = & \int\limits_{-\infty}^{\infty}d\tau_1 \: e^{(\imath\omega + \frac{g}{2} )\tau_1} e^{\frac{\imath p e^{g\tau_1}}{g}}  \nonumber \\
& = & \frac{1}{g}\left ( \frac{g}{p}e^{\frac{\imath\pi}{2}} \right )^{\frac{\imath \omega}{g} + \frac{1}{2}} \Gamma \left (\frac{\imath \omega}{g} + \frac{1}{2} \right )
\end{eqnarray}
as was found in \cite{Dolby3}. If we now calculate $(\beta\beta^\dagger )_{\omega\omega} $, divide it by the normalisation $L = \int\limits_{-\infty}^{\infty} d\rho = \int\limits_{0}^{\infty} \frac{du}{gu}  $, and use the fact that $\vert \Gamma ( \frac{1}{2} + iy ) \vert = \sqrt{\pi \ \rm{sech} \pi y } $ we have
\begin{eqnarray}
\frac{N^+_{\omega, F}}{L} & = & \frac{1}{1 + e^\frac{2\pi\omega}{g}} \nonumber \\
\frac{N^+_{\omega, B}}{L} & = & \frac{1}{1 + e^\frac{2\pi\omega}{g}}
\end{eqnarray}
as found in \cite{Un}, \cite{Dolby3}, and indeed most papers on the subject. 

By calculating $(\beta\beta^\dagger )_{\omega\omega^\prime} $, we find that for $\omega \ne \omega^\prime $ the matrix elements are zero. Hence, since $\psi^\dagger_\omega M \psi_\omega $ is independent of $\rho $, the distribution in space of massless fermions is uniform for this observer.

\subsubsection{Constant Acceleration beginning at $t = 0 $}
\label{conacc}

We can use the results from section \ref{interesting} for massless fermions to integrate over a hypersurface for any value of $\tau $. The simplest choice is $\tau = 0 $ since we need consider only two regions, and 
\begin{eqnarray}
\beta_{\omega \!p,\, F} & = & - \int\limits_{\infty}^{0}d\tau_2 \: e^{(\imath\omega - \frac{g}{2} )\tau_2} e^{\frac{-\imath p e^{-g\tau_2}}{g}}  - \int\limits_{0}^{-\infty}d\tau_2 \: e^{\imath\omega\tau_2 } e^{\imath p (\tau_2 - \frac{1}{g} ) }  \nonumber \\
& = & - \left ( \int\limits_{0}^{\frac{1}{g}} du \: (gu)^{\frac{-\imath\omega}{g} - \frac{1}{2} } e^{\imath p u } \right )  + e^{\frac{-\imath p}{g}} \left ( \pi \delta (\omega + p ) - \frac{\imath}{\omega + p} \right )  \\
& = & \frac{1}{g} \left ( \frac{p}{g} e^{\frac{\imath \pi}{2}} \right )^{\frac{\imath\omega}{g} - \frac{1}{2} } \left [ \Gamma \left ( \frac{-\imath\omega}{g} + \frac{1}{2} \right ) - \Gamma \left ( \frac{-\imath\omega}{g} + \frac{1}{2}, \imath p \right ) \right ] - \imath\frac{e^{\frac{-\imath p}{g}}}{\omega + p} \nonumber 
\end{eqnarray} 
and also
\begin{eqnarray}
\beta_{\omega \!p,\, B} & = &  \int\limits_{-\infty}^{0}d\tau_1 \: e^{\imath\omega\tau_1 } e^{\imath p (\tau_1 + \frac{1}{g} ) }  + \int\limits_{0}^{\infty} d\tau_1 \: e^{(\imath\omega + \frac{g}{2} )\tau_1} e^{\frac{\imath p e^{g\tau_1}}{g}}  \nonumber \\
& = & - \frac{1}{g} \left ( \frac{g}{p} e^{\frac{\imath \pi}{2}} \right )^{\frac{\imath\omega}{g} + \frac{1}{2} } \Gamma\left ( \frac{\imath\omega}{g} + \frac{1}{2}, -\imath p \right ) + \imath\frac{e^{\frac{\imath p}{g}}}{\omega + p}
\end{eqnarray}
where the delta functions are superfluous because $\omega $ must always be positive. The expressions differ for forward- and backward-moving particles; the time symmetry of the constant acceleration case has been broken, so that we no longer expect the numbers of particles moving forwards and backwards to be equal. 

\subsubsection{Constant Acceleration for a finite time}
\label{nohorizon}

Here we use the results from section \ref{turnaround}, and insert them into the massless fermion equations as in the previous examples. As can be seen from Figure \ref{fig:radar3}, the simplest hypersurface of simultaneity to choose for the integration is clearly $\tau = 0$, which corresponds to the $z$-axis and crosses only three regions. Moreover, in regions `IA' and `IB' the coordinates are 
a simple scaling from the inertial coordinates, and the coordinates in region `C' are simply those of the constantly accelerated observer.  We therefore obtain

\begin{eqnarray}
\beta_{\omega \!p,\, F} & = & \left ( \frac{1+v}{1-v} \right )^{\frac{1}{4}} \int\limits_{-\infty}^{-\tau_c} d\tau_2 \: e^{\imath\omega\tau_2}e^{\imath p \sqrt{\frac{1+v}{1-v}}(\tau_2 + \tau_c - \frac{1}{g} )}  \nonumber \\
& & + \int\limits_{-\tau_c}^{\tau_c} d\tau_2 \: e^{(\imath\omega - \frac{g}{2})\tau_2 } e^{\frac{-\imath p e^{-g\tau_2}}{g}} \nonumber \\
& & + \left ( \frac{1-v}{1+v} \right )^{\frac{1}{4}} \int\limits_{\tau_c}^{\infty}d\tau_2 \: e^{\imath\omega\tau_2} e^{\imath p \sqrt{\frac{1-v}{1+v}}(\tau_2 - \tau_c - \frac{1}{g})} 
\end{eqnarray}
and 
\begin{eqnarray}
\beta_{\omega \!p,\, B} & = & \left ( \frac{1-v}{1+v} \right )^{\frac{1}{4}} \int\limits_{-\infty}^{-\tau_c} d\tau_1 \: e^{\imath\omega\tau_1}e^{\imath p \sqrt{\frac{1+v}{1-v}}(\tau_1 + \tau_c + \frac{1}{g} )}  \nonumber \\
& & + \int\limits_{-\tau_c}^{\tau_c} d\tau_1 \: e^{(\imath\omega + \frac{g}{2})\tau_1 } e^{\frac{\imath p e^{g\tau_1}}{g}}\nonumber \\
& & + \left ( \frac{1+v}{1-v} \right )^{\frac{1}{4}} \int\limits_{\tau_c}^{\infty} d\tau_1 \: e^{\imath\omega\tau_1} e^{\imath p \sqrt{\frac{1-v}{1+v}}(\tau_1 - \tau_c + \frac{1}{g})} \nonumber \\
& = & (\beta_{\omega \!p,\, F})^*
\end{eqnarray}

From these results, we can determine the particle content of the vacuum according to the observer, despite the absence of an acceleration horizon. The results illuminate the earlier proof that for a time-symmetric observer worldline the `forward'- and `backward'-moving particles have the same number and the same frequency distribution.

\section{Discussion}

The approach developed in \cite{Dolby3} has been used to evaluate the particle content of the massless Dirac vacuum for an arbitrarily moving observer in 1+1 dimensions. This method uses Bondi's `radar time', which provides a foliation of flat space that depends only on the observer's trajectory. It agrees with proper time on the trajectory, and is single valued in the causal envelope of the observer's worldline. 

The particle content can be described in terms of `forward moving' particles (with particle density $n^+_F(\tau,\rho) $ dependent only on $\tau - \rho$), `backward moving' particles (with particle density
$n^+_B(\tau,\rho)$ dependent only on $\tau + \rho$), and their corresponding antiparticles (described by $n^-_F$ and $n^-_B $). The particle content is clearly different for different observers, but for any given observer the total number of forward/backward moving particles is independent of $\tau$. This result is specific to massless particles in 1+1 dimensions, and is a direct consequence of conformal invariance. The spatial distribution of forward moving particles always matches that of forward moving antiparticles, and likewise for backward moving particles. In general, there is no connection between the observed numbers of forward moving particles and backward moving particles. However, when the observer's trajectory is time-reversal invariant the distribution of forward and backward moving particles must match at $\tau=0 $, and their distributions at other times is given by the fact that $n^{\pm}_F(\tau,\rho) $ depends only on $\tau - \rho $ and that $n^{\pm}_B(\tau,\rho) $ depends only on $\tau + \rho$. These general results have been illustrated by four examples. We have confirmed that inertial observers do not observe particles in the inertial vacuum, and that uniformly accelerating observers measure the well-known thermal spectrum [1,2,3,8,12,13]. This thermal spectrum is spatially uniform in `radar distance' $\rho $ and is the same for forward and backward moving particles, as expected. An observer with constant acceleration for finite time (who is otherwise inertial) exhibits time-reversal invariance about the time ($\tau=0 $ in section \ref{nohorizon}) mid way through the period of acceleration. This gives $n^{\pm}_F(0,\rho) = n^{\pm}_B(0,\rho)$ as predicted, although these distributions are no longer spatially uniform. An observer who accelerates uniformly from $\tau=0 $ observes different numbers of forward and backward moving particles. This intriguing result suggests an observer-dependent violation of discrete symmetries, which warrants further investigation. 

Other work that has treated non-uniformly accelerating observers in flat space is presented in \cite{Yang}. There, the content of massive scalar particles was studied in flat space, using a foliation based on the observer's `instantaneous rest frame'. Interesting results were obtained there, but their interpretation is limited by the multivalued nature of the instantaneous rest frame, as discussed in \cite{Dolbyradar}. The use of an `instantaneous rest frame' will in general assign more than one time to any event further than $1/g(t) $ from an accelerating observer (uniform acceleration is the only exception). This leads in \cite{Yang} to multivalued predictions for the particle content at any such event, raising problems of interpretation.

Since massless fermions in 1+1 dimensions are conformally invariant, and consequently lack the scale dependence of the massive case(provided for instance by the Compton scale $1/m $), it is not likely that the results obtained here will generalise immediately to massive particles, or to 3+1 dimensions. Further work is needed to investigate those cases in more detail, and to explore the possible observer dependence of discrete symmetries.

\newpage
\appendix
\section{Identities}

This Appendix sets out many identities used here, mostly following from simple partial derivatives.

\begin{eqnarray}
\frac{\partial \tau}{\partial x^+} & = & \frac{1}{2} \frac{\partial }{\partial x^+} \left ( \tau_1 + \tau_2 \right ) \nonumber \\
& = & \frac{1}{2} \frac{d\tau_1}{dx^+} \nonumber \\
& = & \frac{1}{2U^+}
\end{eqnarray}
and similarly
\begin{equation}
\frac{\partial \tau}{\partial x^-} = \frac{1}{2U^-}
\end{equation}

From our radar coordinate definitions $\tau = \frac{1}{2} (\tau_1 + \tau_2 ) $ and $\rho = \frac{1}{2} (\tau_1 - \tau_2 ) $, we have
\begin{equation}
\frac{\partial \tau}{\partial x^+ } = \frac{\partial \rho}{\partial x^+}, \qquad \frac{\partial \tau}{\partial x^- } = - \frac{\partial \rho}{\partial x^- }
\end{equation}
and
\begin{equation}
\frac{\partial \tau}{\partial t} = \frac{\partial \rho}{\partial z}, \qquad \frac{\partial \tau}{\partial z} = \frac{\partial \rho}{\partial t}
\end{equation}

From the identities $\rho = \tau - \tau_2 = \tau_1 - \tau $ we can obtain
\begin{equation}
\left ( \frac{\partial}{\partial \rho} \right )_\tau  =  \left ( \frac{\partial}{\partial \tau_1} \right )_\tau = - \left ( \frac{\partial}{\partial \tau_2} \right )_\tau
\label{rhoparams} \end{equation}

Identities involving the $a $, $b $ and $\sigma $ matrices of section \ref{Hamiltonian} are
\begin{eqnarray}
b_1^2 = 2 b_1, \qquad b_2^2 = 2 b_2 \nonumber \\
b_1 b_2 = b_2 b_1 = 0 \nonumber \\
b_1 + b_2 = \mathbb{I}, \qquad b_1 - b_2 = 2\sigma
\end{eqnarray}
and
\begin{eqnarray}
a_1 b_1 = 2a_1, \qquad a_1 b_2 = 0 \nonumber \\
b_1 a_1 = 0, \qquad b_2 a_1 = 2a_1 \nonumber \\
a_2 b_1 = 0, \qquad a_2 b_2 = 2a_2 \nonumber \\
b_1 a_2 = 2a_2, \qquad b_2 a_2 = 0
\end{eqnarray}
together with the commutators:
\begin{eqnarray}
\lbrack b_1, b_2 \rbrack = 0, \qquad \lbrack a_1, a_2 \rbrack = -4 \sigma   \nonumber \\
\lbrack \sigma , b_1 \rbrack = \lbrack \sigma , b_2 \rbrack = 0   \nonumber \\
\lbrack \sigma , a_1 \rbrack = -2 a_1 , \qquad \lbrack \sigma , a_2 \rbrack = -2 a_2  \nonumber \\
\label{commutators} \end{eqnarray}

\end{document}